# Control of resonant frequency by currents in graphene: Effect of Dirac field on deflection

Bumned Soodchomshom[a,*]

[a] Department of Physics, Faculty of Science, Kasetsart University Bangkok 10900, Thailand

*Corresponding author
Email address: Bumned@hotmail.com; fscibns@ku.ac.th

**Abstract**

To construct Lagrangian based on plate theory and tight-binding model, deflection-field coupling to Dirac fermions in graphene can be investigated. As have been known, deflection-induced strain may cause an effect on motion of electron, like a pseudo gauge field. In the work, we will investigate the effect of the Dirac field on the motion of the deflection-field in graphene derived from Lagrangian density. Due to the interaction of the deflection- and Dirac-fields, the current-induced surface-tension up to about $4 \times 10^{-3}$ N/m in graphene membrane is predicted. This result may lead to controllable resonant frequency by currents in graphene. The high resonant frequency is found to be perfectly linearly controlled by both charge and valley currents. Our work reveals the potential of graphene for application of nano-electro-mechanical device and the physics of interaction of electron and deflection-filed in graphene system is investigated.

**1. Introduction**

After the discovery of graphene, a stable monolayer of graphite [1], graphene has become one of promising materials for application of nano-electro-mechanical devices [2-5]. Graphene has large strain sustaining [6-7] which allows it to be a good flexible and stretchable nano-electronics [8-12]. The Young's modulus of graphene was found approximately about 1 TPa [7]. Recently, both experimental [2, 13-20] and theoretical [21-26] investigations of resonant frequency mode of suspended graphene have drawn much attention. The vibrational-bending mode of graphene due to its elasticity leads to the interesting application of a resonator [2, 15, 18, 19] and mass sensing devices [21-25]. The plate theory was first adopted to study the interaction of electron-phonon scattering in carbon nanotubes [27]. The motion of electrons under the effect of deflection and displacement fields was studied. The plate theory [27] has been next extensively used to investigate the interplay of bending of graphene and Dirac fermionic field to create a pseudo gauge field in graphene [28, 29]. As have been known, electrons in graphene feel the bending of graphene atomic structure as a pseudo gauge potential [30-31]. A gigantic strain-induced pseudo magnetic field greater than 300 Tesla in graphene nanobubble was reported experimentally [32].

One of the interesting electronic properties of graphene is that, honeycomb like atomic structure made up of the carbon atoms, which has two sublattics A and B in a unit cell, allows its electrons to behave like the massless relativistic particle with two valley degrees of freedoms k- and -valleys [28, 29]. The wave states in the A- and B-sublatices have been considered as equivalent to pseudo spin-up and -down states, respectively. Under the influence of bending (or strain) of graphene, two-valley electrons may feel the pseudo magnetic field with different signs. In previous works [28, 29], the investigation was to focus on how the deflection and displacement in graphene affect on electron field in graphene. The effect of the Dirac electron field on the bending of graphene has not been clarified.

In this paper, we would investigate the effect of the Dirac field which is obtained from the tight-binding model [30, 31] on the bending of graphene. The Lagrangian density of free bending graphene is constructed based on the plate theory [33]. The combination of the plate and tight-binding model may lead to construction of interacting Lagrangian density of bending plate and the Dirac field. We focus on the effect of Dirac field on the resonant mode of out-of-plane graphene vibration or the deflection-filed derived from the Lagrangian density. Using this methodology, we will show that the resonant frequency of graphene vibration may be controlled by currents of the junction. The current density of the Dirac field behaves like anisotropic stress to change resonant frequency mode in a graphene-based resonator.

**2. Lagrangian density of the interaction of Dirac- and deflection-fields**

To first consider geometry of graphene atomic structure depicted in Fig.1(a), the x (y) direction is assumed to be parallel to the zigzag (armchair) direction. The deflection $u_z$ is the out-of-plane displacement of graphene. The in-plane displacements along the x-and y- directions are $u_x$ and $u_y$ respectively. The pseudo gauge potentials $A_x, A_y$ and scalar potential $\phi$ created by the changes of $<u_x, u_y, u_z>$ are given respectively by [27-31]

$$A_x = \frac{g_2}{ev_F}(u_{xx} - u_{yy}), \quad A_y = -2\frac{g_2}{ev_F}u_{xy} \text{ and } \phi = \frac{g_1}{e}(u_{xx} + u_{yy}),$$

where
$$u_{xx} = \frac{\partial u_x}{\partial x} + \frac{1}{2}\left(\frac{\partial u_z}{\partial x}\right)^2, \quad u_{yy} = \frac{\partial u_y}{\partial y} + \frac{1}{2}\left(\frac{\partial u_z}{\partial y}\right)^2$$

and
$$u_{xy} = \frac{1}{2}\left(\frac{\partial u_x}{\partial x} + \frac{\partial u_y}{\partial y}\right) + \frac{1}{2}\frac{\partial u_z}{\partial x}\frac{\partial u_z}{\partial y}.$$

(1)

$g_1$ and $g_2$ are the coupling constants, e is the charge of bare electron and $v_F = 10^6$ m/s is the Fermi velocity of electron in graphene. In our focus, the effect of the in-plane displacements on the motion of Dirac electron filed is neglected. The pseudo gauge potential therefore depends only on the transverse displacement, the deflection-field $u_z = u_z(x, y, t)$. By doing this, the pseudo gauge and scalar potentials may be respectively given as of the forms

$$A_x = \frac{g_2}{2ev_F}\left(\left(\frac{\partial u_z}{\partial x}\right)^2 - \left(\frac{\partial u_z}{\partial y}\right)^2\right), \quad A_y = -\frac{g_2}{ev_F}\frac{\partial u_z}{\partial x}\frac{\partial u_z}{\partial y}$$

$$\text{and } , \phi = \frac{g_1}{2e}\left(\left(\frac{\partial u_z}{\partial x}\right)^2 + \left(\frac{\partial u_z}{\partial y}\right)^2\right).$$

(2)

The motions of the valley-dependent Dirac field $\psi_{k(k')} = \psi_{k(k')}(x,y,t)$ and the deflection-filed, $u_z(x,y,t)$, in graphene may be described using the Lagrangian density below

$$L = L_{Free-electron} + L_{Free-deflection} + L_{interaction}. \qquad (3)$$

In eq.(3), the Lagrangian density of the free-Dirac-field of electron in graphene may be written as

$$L_{Free-electron} = i\hbar v_F \sum_{\xi=k,k'} \left( \psi_\xi^* \frac{\partial}{v_F \partial t}\psi_\xi + \psi_\xi^* \sigma_x \frac{\partial \psi_\xi}{\partial x} + \psi_\xi^* \sigma_y \frac{\partial \psi_\xi}{\partial y} \right), \qquad (4)$$

where $\xi = k$ and $k'$ and $\sigma_{x,y}$ are Pauli matrices acting on the lattice pseudo spin wave states. Based on the plate theory [33], the Lagrangian density of the free deflection field or the out-of-plane displacement may be given as

$$L_{Free-deflection} = \frac{\rho_A}{2}\left(\frac{\partial u_z}{\partial t}\right)^2 - \frac{\kappa}{2}\left(\nabla^2 u_z\right)^2, \qquad (5)$$

where $\rho_A = 7.6\times 10^{-7}\,kg/m^2$ is the mass density of graphene per unit area and $\nabla^2 = \partial^2/\partial x^2 + \partial^2/\partial y^2$ is the two-dimensional Laplacian. The plate bending stiffness is defined as $\kappa = Yh^3/12(1-\nu^2)$, where $Y$, $h$ and $\nu$ are the Young's modulus, the thickness and the Poisson's ratio of graphene, respectively. Based on tight-biding theory [30, 31], the valley-dependent interaction between the Dirac- and deflection-fields may be written as

$$L_{interaction} = ev_F \sum_{\xi=k,k'} \left( -\psi_\xi^*(\frac{\phi}{v_F})\psi_\xi + \eta_\xi \psi_\xi^* \sigma_x A_x \psi_\xi + \eta_\xi \psi_\xi^* \sigma_y A_y \psi_\xi \right), \qquad (6)$$

where $\eta_{\xi=k} = -\eta_{\xi=k'} = -1$. Here, $A_{x,y}$ and $\phi$ in eq.(6) are based on formulae in eq.(2).

**3. Motion of the deflection-field under the influence of the Dirac field**

The action of the system is defined as $S = \iint L dx dy dt$. The equations of motions for electron and the deflection may be calculated via the variation of the action

$$\delta S = 0. \tag{7}$$

As we have seen in eqs.(3-6) that the Lagrangian density contains derivatives with respect to x and y higher than the first, such problem has been referred to as "Jerky mechanics" [34]. Hence, the corresponding Euler–Lagrange equations for electron-field $\psi_\xi$ and the deflection-field $u_z$ may be respectively given as

$$\sum_{i=1}^{3}\left\{\frac{\partial}{\partial x_i}\left(\frac{\partial L}{\partial\left(\partial\psi_\xi^*/\partial x_i\right)}\right)-\frac{\partial^2}{\partial x_i^2}\left(\frac{\partial L}{\partial\left(\partial^2\psi_\xi^*/\partial x_i^2\right)}\right)\right\}=\frac{\partial L}{\partial\psi_\xi^*} \tag{8}$$

and

$$\sum_{i=1}^{3}\left\{\frac{\partial}{\partial x_i}\left(\frac{\partial L}{\partial\left(\partial u_z/\partial x_i\right)}\right)-\frac{\partial^2}{\partial x_i^2}\left(\frac{\partial L}{\partial\left(\partial^2 u_z/\partial x_i^2\right)}\right)\right\}=\frac{\partial L}{\partial u_z}, \tag{9}$$

where $x_1 = x$, $x_2 = y$ and $x_3 = t$.

From eq.(8), we get equations of motion for electron filed $\psi_\xi$ for $\xi = k, k'$-valleys of the forms

$$i\hbar\frac{\partial\psi_\xi}{\partial t}=\left\{v_F[-i\hbar\frac{\partial}{\partial x}-\eta_\xi eA_x]\sigma_x + v_F[-i\hbar\frac{\partial}{\partial y}-\eta_\xi eA_y]\sigma_y + e\phi\right\}\psi_\xi. \tag{10}$$

This equation is known as Dirac equation under pseudo gauge and scalar potentials when graphene is under strain filed induced by the deflection [31]. The electrons in k- and k'-valleys feel opposite pseudo gauge potential but the same for scalar potential [27, 31]. In case of the equation of motion for the deflection-filed which may be derived from eq.(9), we get

$$\rho_A\frac{\partial^2 u_z}{\partial t^2}+\kappa\nabla^4 u_z = T_{xx}\frac{\partial^2 u_z}{\partial x^2}+T_{yy}\frac{\partial^2 u_z}{\partial y^2}+T_{xy}\frac{\partial^2 u_z}{\partial x\partial y}+T_{yx}\frac{\partial^2 u_z}{\partial y\partial x}, \tag{11}$$

where, here, the wave function of electron is assumed to be plane wave state, leading to that $\frac{\partial\psi_\xi^*\psi_\xi}{\partial x(y)}=\frac{\partial\psi_\xi^*\sigma_x\psi_\xi}{\partial x(y)}=\frac{\partial\psi_\xi^*\sigma_y\psi_\xi}{\partial x(y)}=0$. We thus get

$$T_{xx} = g_1\left(\psi_{k'}^*\psi_{k'} + \psi_k^*\psi_k\right) + g_2\left(\psi_k^*\sigma_x\psi_k - \psi_{k'}^*\sigma_x\psi_{k'}\right),$$

$$T_{yy} = g_1\left(\psi_{k'}^*\psi_{k'} + \psi_k^*\psi_k\right) - g_2\left(\psi_k^*\sigma_x\psi_k - \psi_{k'}^*\sigma_x\psi_{k'}\right),$$

and $\quad T_{xy} = T_{yx} = g_2\left(\psi_{k'}^*\sigma_y\psi_{k'} - \psi_k^*\sigma_y\psi_k\right).$

(12)

Interestingly, as seen in eqs.(11) and (12), it can be considered that the Dirac electron fields may cause the stress-tensor $\tau_{\mu\nu}$ into graphene membrane (case of $T_{\mu\nu}$ =constant with $\mu,\nu = \{x,y\}$), given as

$$\tau_{\mu\nu} = \frac{1}{h}\begin{bmatrix} T_{xx} & T_{xy} \\ T_{yx} & T_{yy} \end{bmatrix}, \quad (13)$$

into graphene membrane as an external anisotropic source $f_{source}$ of the form

$$f_{source} = T_{xx}\frac{\partial^2 u_z}{\partial x^2} + T_{yy}\frac{\partial^2 u_z}{\partial y^2} + T_{xy}\frac{\partial^2 u_z}{\partial x \partial y} + T_{yx}\frac{\partial^2 u_z}{\partial y \partial x}. \quad (14)$$

The source term mimics the behavior of anisotropic membrane vibration. The free out-of-plane vibration of thin plate is, for $f_{source} = 0$, given in the form [33]

$$\rho_A \frac{\partial^2 u_z}{\partial t^2} + \kappa \nabla^4 u_z = 0 \quad (15)$$

This result may give rise to the control of vibration mode in graphene by electronic property which may be very important for application of nano-electro-mechanical device. The mechanical vibration property may be controlled by electronic filed.

**4. Current density dependence of graphene vibration**

The wave function of electron in graphene $\psi_\xi$ is two-dimensional-pseudo spinor field. The valley-current density $\vec{J}_\xi$ of electrons obeying two-dimensional-Dirac equation given in eq.(10) may be determined through the standard continuity condition

$$\vec{\nabla}\cdot\vec{J}_\xi = -\frac{\partial \rho_{e,\xi}}{\partial t}, \quad (16)$$

where $\vec{J}_\xi = \frac{ev_F}{h}\left[\psi_\xi^*\sigma_x\psi_\xi \hat{i} + \psi_\xi^*\sigma_y\psi_\xi \hat{j}\right]$ and $\rho_{e,\xi} = \frac{e}{h}\psi_\xi^*\psi_\xi$ is valley-charge-density. The valley-current density in the x(y) direction is

$$J_{x(y),\xi} = \frac{ev_F}{h}\psi_\xi^*\sigma_{x(y)}\psi_\xi. \qquad (17)$$

The norm of the total valley-current-density may be obtained using the formula

$$\left|\vec{J}_\xi\right| = \sqrt{\left(\frac{ev_F}{h}\psi_\xi^*\sigma_x\psi_\xi\right)^2 + \left(\frac{ev_F}{h}\psi_\xi^*\sigma_y\psi_\xi\right)^2} = \frac{e\sqrt{2}v_F}{h}\psi_\xi^*\psi_\xi = J_\xi.$$

(18)

By substituting eqs.(17) and (18) into eq.(12), we get

$$T_{xx} = \tilde{g}_1(J_k + J_{k'}) + \tilde{g}_2(J_{x,k} - J_{x,k'}),$$

$$T_{yy} = \tilde{g}_1(J_k + J_{k'}) - \tilde{g}_2(J_{x,k} - J_{x,k'}),$$

and $\quad T_{xy} = T_{yx} = \tilde{g}_2(J_{y,k'} - J_{y,k}),$

where $\tilde{g}_1 = \frac{hg_1}{ev_F\sqrt{2}}$ and $\tilde{g}_2 = \frac{hg_2}{ev_F}$.

(19)

The total current density may be defined as of the form $J = J_{k'} + J_k$, since the current of the junction is carried by the two groups, k- and k'-currents. In Eq.(19), it is also shown that the vibration of graphene may be directly influenced by the current of the junction. This result should be significant for application of current-controlled resonant frequency in a graphene-based resonator.

**5. Possible current-controlled resonant frequency in graphene-based electronic junction**

In this section, the electronic model in Fig.1(b) to study the current-controlled vibration is theoretically investigated. Suspended graphene sheet with length "L" and width "W" is modeled as a doubly-clapped beam. The current density J is assumed to flow in the x-direction. In our limit, we assume that there is no vibration mode along the y-direction. Hence, the deflection-filed may depend only on x and t, ie., $u_z = \chi(x)T(t)$. The graphene plate, therefore, can be considered as a doubly-clapped beam. The current density is flow only in the x-direction then $J_{x,k} = J_k$ and $J_{x,k'} = J_{k'}$. The motion of the deflection-field in graphene-based doubly-clamped beam may be described by reducing eq.(11) where we take $\frac{\partial^2 u_z}{\partial x \partial y} = \frac{\partial^2 u_z}{\partial y \partial x} = \frac{\partial^2 u_z}{\partial y^2} \Rightarrow 0$. The equation of motion for the deflection is thus given by

$$\rho_A \frac{\partial^2 u_z}{\partial t^2} + \kappa \frac{\partial^4 u_z}{\partial x^4} - (T_{current} + T_o) \frac{\partial^2 u_z}{\partial x^2} = 0,$$

(20)

where $T_{xx} \to T_{current} = (\tilde{g}_1 + \tilde{g}_2 p_{valley}) \times J$ is current-induced surface tension with $p_{valley} = (J_k - J_{k'})/J$ being valley polarization of the junction. We note that $T_o$ is a possible added-initial surface tension, due to fabricating junction. As a result above, our theory shows that the vibration mode of suspended graphene in our electronic junction may be described using eq.(20). The deflection $u_z$ has been predicted that it can be controlled by the biased current "J", because $T_{current}$ depends directly on biased current passing through suspended graphene. Hence, varying current of the junction would change the resonant frequency of a graphene vibrator, modeled in Fig.1(b). The calculation of current dependence of resonant frequency will be next studied.

The solution to eq.(20) may be calculated by assuming the solution of the form $u_z = \chi(x) e^{-i\omega t}$ where $f = \omega / 2\pi$ is vibration frequency. Hence, the time-independent solution can be written as

$$\chi(x) = c_1 \sin(\Omega_- x) + c_2 \cos(\Omega_- x) + c_3 \sinh(\Omega_+ x) + c_4 \cosh(\Omega_+ x),$$

(21)

where $\Omega_\pm = \left[ \sqrt{(T_{current} + T_0)^2 + 4\kappa \rho_A \omega^2} \pm (T_{current} + T_0) \right]^{1/2} / \sqrt{2\kappa}$ and $c_{1,2,3,4}$ are amplitude coefficients. The boundary conditions for doubly-clamped beam (vanishing of force and moment) is usually given as

$$\chi(0) = \frac{\partial \chi(x)}{\partial x}\bigg|_{x=0} = 0 \text{ and } \chi(L) = \frac{\partial \chi(x)}{\partial x}\bigg|_{x=L} = 0.$$

(22)

By matching the time-independent solution with the boundary condition, the resonant frequency may be calculated via a non-zero solution condition to get

$$\left[\Omega_+ \sin(\Omega_- L) - \Omega_- \sinh(\Omega_+ L)\right]\left[\Omega_- \sin(\Omega_- L) + \Omega_+ \sinh(\Omega_+ L)\right] +$$
$$\Omega_+ \Omega_- \left[\cos(\Omega_- L) - \cosh(\Omega_+ L)\right]^2 = 0.$$

(23)

In numerical calculation of current density dependence of resonant frequency, the thickness of graphene $h = 3.35\,\overset{o}{A}$ is approximated [35]. The elastic constants, Young's modulus $Y = 1\,\text{Tpa}$ [7], Poisson ratio $\nu = 0.17$ [35] may give rise to the plate bending stiffness of about $\kappa = 20\,\text{eV}$. The coupling constants of interaction of electron- and deflection-fields are set to be of about $g_1 = 16\,\text{eV}$ [27] and $g_2 = 2.3\,\text{eV}$ [29] at the temperature 300 K. By using the above constants, the current-induced surface tension may be obtained as

$$T_{current} = \left(5.4\times10^{-15} + 0.8\times10^{-15} p_{valley}\right)\times J \quad \text{(N/m)}. \quad (24)$$

Since the magnitude of breakdown current density $J_{break} = 10^{12}\ \text{A}/\text{m}^2$ in graphene was adopted from the experimental report in ref.[36], this breakdown current density would lead to the maximum current-induced surface tension for no valley-polarized current ($p_{valley} = 0$) of about $T_{Max-current} = 3.8\times10^{-3}\,\text{N/m}$. The possible magnitude of current density in graphene junction studied in this work is assumed below $J_{break}$.

In our calculation, we set the studied junction with zero initial surface tension $T_0 = 0$. The contour plot of resonant frequency "f" versus the current density "J" calculated using eq.(23) is first studied in Figs.2(a) and 2(b). Interestingly, the resonant frequency is found to be perfectly linearly controlled by the current density for which the length of the suspended graphene is small. In the regime of high frequency for case of $L = 100\,\text{nm}$, it is found that the fundamental frequency (n=1) perfectly linearly depends on J, varying from $f_{min} = 730\,\text{MHz}$ to $f_{max} = 830\,\text{MHz}$. In Figs.3(a) and 3(b), it is also shown that the resonant frequency may depend perfectly linearly on the valley-polarization despite large L. The effect of current density on fundamental frequency increases strongly as L decreases. The predicted linear response of resonant frequency under varying currents in graphene may be applicable for graphene-based resonator devices in which its resonant frequency is tunable by electric current.

## 6. Summary

We have investigated the interaction of electron or Dirac-field and the deflection-field in graphene based on the Lagrangian density constructed using plate and tight-binding theories. It was found that the current density of electron filed behaves like an external anisotropic source term. Current-induced surface tension in graphene membrane has been predicted. The application of our result was to shown a possible perfect linear control of resonant frequency in graphene-based resonator using the charge current and valley polarization. Our work has revealed the effect of current density of Dirac field on the vibration of graphene and showed the possible application for nano-electro-mechanical devices using graphene-based material.

**Acknowledgments**

This work was supported by Kasetsart University Research and Development Institute (KURDI) and Thailand Research Fund (TRF) under Grant. No.TRG5780274.

**Figure captions**

**Figure 1** Schematic illustrations of (a) graphene atomic structure with two sub-lattices A and B in a unit cell and (b) a doubly-clamped beam model using suspended graphene with length L and width W. The x(y) direction is parallel to the zigzag (armchair) direction. The current density J is assumed to flow only in the x-direction.

**Figure 2** Contour plot of the resonant frequency versus the current density of the junction (a) $L = 1000$ nm (b) $L = 100$ nm. In this case, it is assumed to have no valley-polarization and no initial surface tension across the junction.

**Figure 3** Contour plot of resonant frequency versus the valley-polarization of the junction for (a) $L = 1000$ nm (b) $L = 100$ nm. In this case, it is assumed to have no initial surface tension across the junction.

(a)

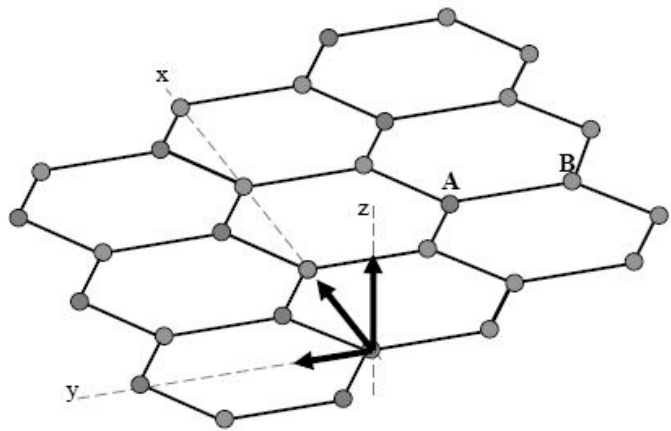

(b)

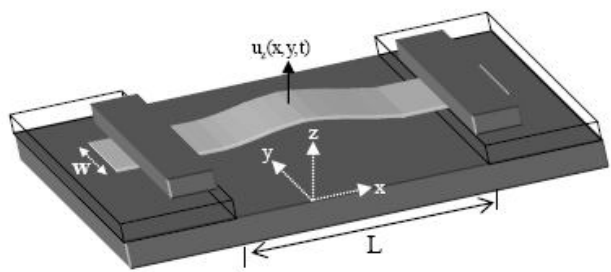

**Figure 1**

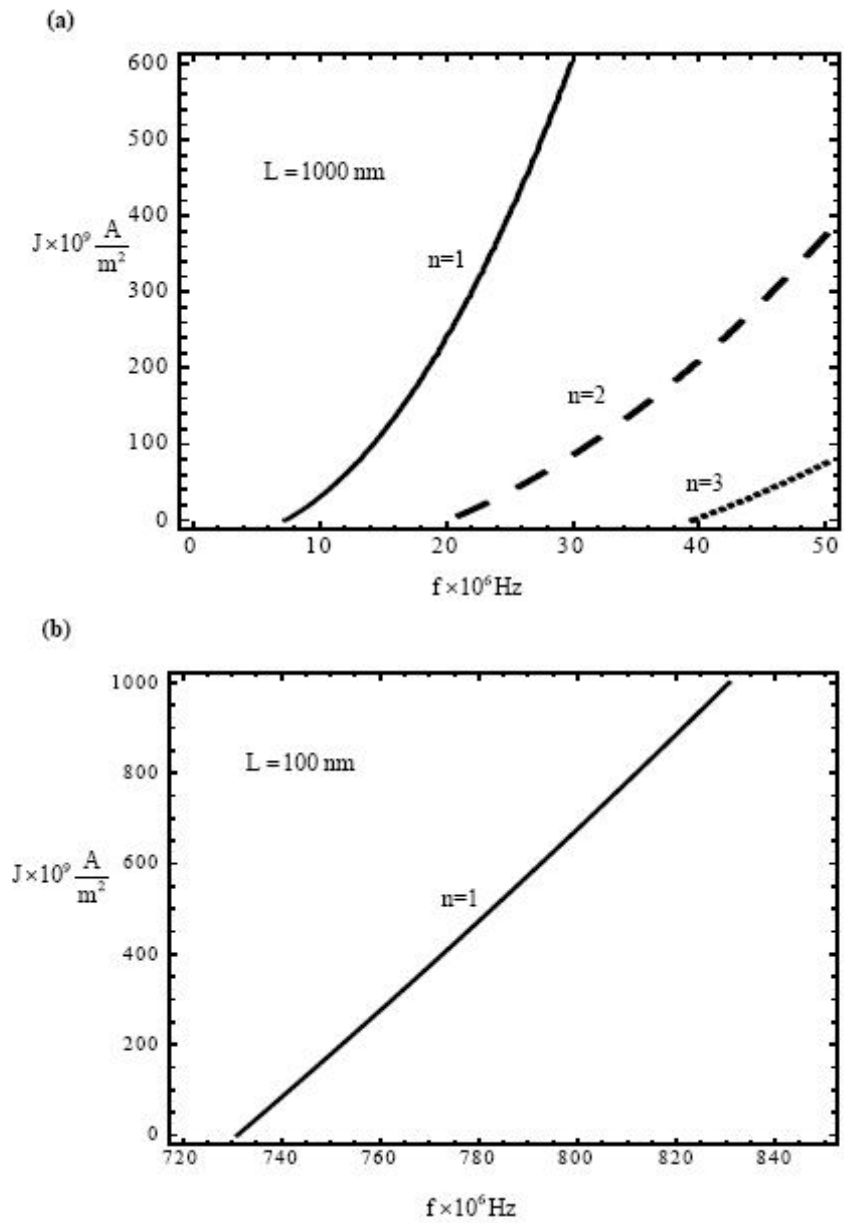

**Figure 2**

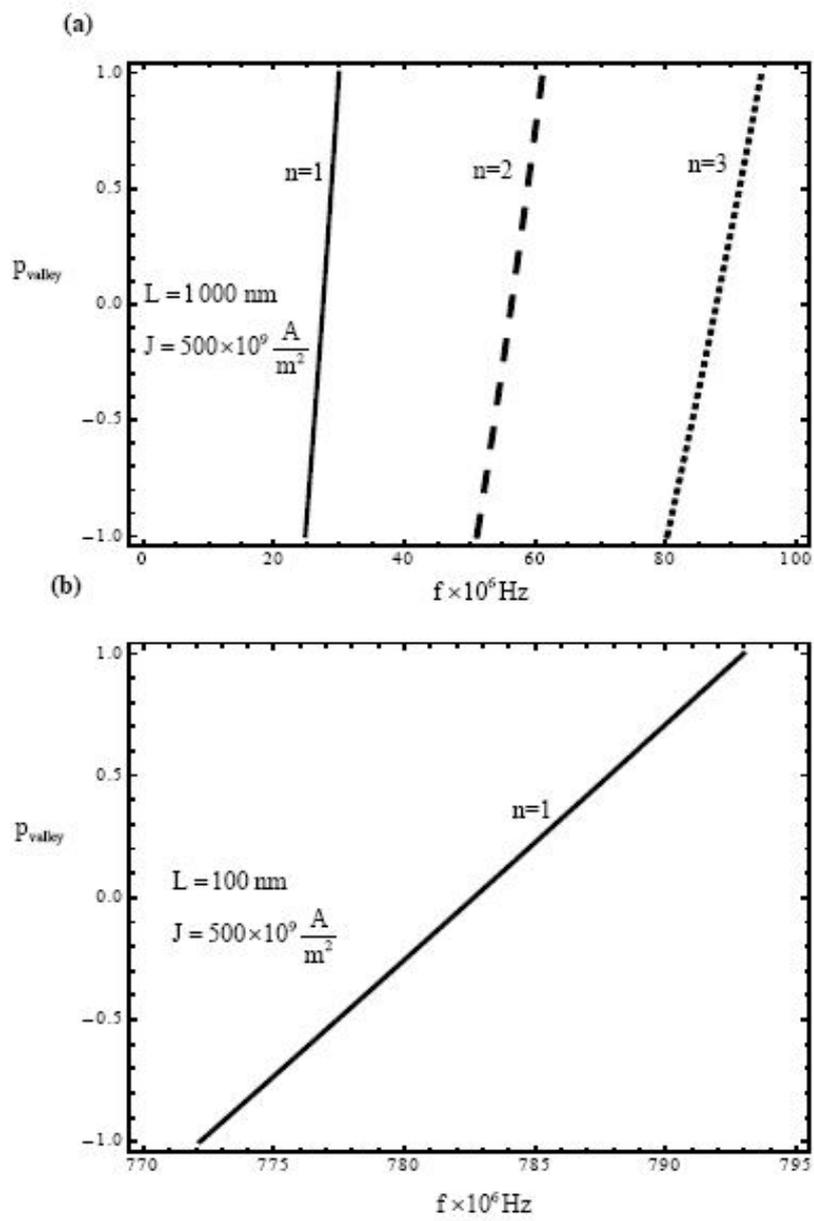

**Figure 3**